# Subamorphous thermal conductivity of crystalline half-Heusler superlattices

E. Chavez-Angel[1,*,†], N. Reuter[1*], P. Komar [1,2,‡], S. Heinz[1,2], U. Kolb[3,4], H.-J. Kleebe[4] and G. Jakob[1,2]

[1] Institut für Physik, Johannes Gutenberg Universität Mainz, Staudingerweg 7, 55128 Mainz, Germany.
[2] Graduate School Materials Science in Mainz, Staudingerweg 9, 55128 Mainz, Germany
[3] Institute of Inorganic and Analytical Chemistry, Johannes Gutenberg Universität Mainz, Duesbergweg 10-14, 55128 Mainz, Germany.
[4] Institute of Applied Geosciences, TU Darmstadt, Schnittspahnstraße 9, 64287 Darmstadt, Germany.

E-mail corresponding author: emigdio.chavez@icn2.cat

The quest to improve the thermoelectric figure of merit has mainly followed the roadmap of lowering the thermal conductivity while keeping unaltered the power factor of the material. Ideally an electron-crystal phonon-glass system is desired. In this work, we report an extraordinary reduction of the cross-plane thermal conductivity in crystalline (TiNiSn):(HfNiSn) half-Heusler superlattices. We create SLs with thermal conductivities below the effective amorphous limit, which is kept in a large temperature range (120-300 K). We measured thermal conductivity at room temperature values as low as 0.75 W m$^{-1}$ K$^{-1}$, the lowest thermal conductivity value reported so far for half-Heusler compounds. By changing the deposition conditions, we also demonstrate that the thermal conductivity is highly impacted by the way the single segments of the superlattice grow. These findings show a huge potential for thermoelectric generators where an extraordinary reduction of the thermal conductivity is required but without losing the crystal quality of the system.

---

[*] Equally contributing authors
**Current address**
[†] Catalan Institute of Nanoscience and Nanotechnology (ICN2), CSIC and BIST, Campus UAB, Bellaterra, 08193 Barcelona, Spain.
[‡] Photonics Group, Institute of Physics, Lodz University of Technology, Wólczańska 219, 90-924 Łódź, Poland.

## 1. Introduction

Understanding of heat propagation and the ability to tune the thermal properties constitute a topic of continuous and active research motivated by the increasing importance of thermal management and ways to recover waste heat energy as it is the case for the thermoelectric industry. This renewed interest in thermal management has introduced a number of novel concepts and ideas including: thermocrystals[1], thermal-cloaking, -transistors, -diodes and -memories[2–9], phonon-mean-free-path spectroscopy[10,11], among others.

The control of phonon propagation, the main heat carriers in semiconductors and insulators, is a crucial requirement for thermoelectric generation. Ideally, a material with thermal properties of an amorphous state (phonon glass) and electronic properties associated with good single-crystal semiconductor (electron crystal) are desired.[12] Materials very low thermal conductivity, $k$, are also needed in other applications such as: thermal barrier coatings for gas turbine engines and thermal data storage devices.[13]

The lowest $k$ in crystalline systems is achievable through alloy scattering, or the so-called alloy limit. But, the introduction of an extra scattering mechanisms, e.g., nanostructures, can exceed this limit. The use of superlattices[14–17] and embedded nanoparticles[18] have demonstrated to be a good way to reduce $k$ below the alloy limit, while maintaining the crystal quality of the material. The introduction of more and more scattering events can reduce even further this limit reaching its second minima: the amorphous limit. In this context, recent experiments showed that, by introducing small-periods in SLs, ultralow $k$ values below the amorphous limit can be achieved.[19–21] Costescu *et al*.[19] and Pernot *et al*.[20] measured cross plane thermal conductivity values ($k_\perp$) below the amorphous limit of $Al_2O_3$ and Si in $Al_2O_3$:W and SiGe:Si SLs, respectively. Niemelä *et al*. also overtook the amorphous limit of $TiO_2$ using organic-inorganic ($TiO_2$):(Ti–O–$C_6H_4$–O) SLs. Moreover, Chiritescu *et al*.[22] also measured ultralow $k_\perp$ in layered $WSe_2$ thin films. $k_\perp$ values below the amorphous limit of $WSe_2$ crystal was achieved by controlling both order and disorder in the thin films.

In SLs, thus, it is natural to think that the smaller period length ($L$), the smaller $k_\perp$. However, several theoretical[15,23–25] and experimental[17,26,27] reports have shown that for very thin $L$ the $k_\perp$ increases. In such limit, phonons experience the material as if it was composed of enlarged unit cells given by the size of $L$. The SL is seen as one homogenous material and the phonon transport is considered coherent.[28] The transition between coherent-incoherent (wave-particle) transport is observed as a minimum in the thermal conductivity, $k_\perp$ as a function of $L$[17,26]. This effect comes from the competition between phonons diffusively scattered at each interface and the band-folded ones. The first unambiguous experimental demonstration of this crossover was presented by Ravichandran *et al.*[17] using epitaxial perovskite-based SLs. Another fingerprint of coherent thermal transport was proposed by Luckyanova *et al.*[28], namely, a linear dependence of $k_\perp$ with respect to the number of periods as indicator of coherent thermal transport through the SL. This arises when the phonon mean free paths (MFPs) are equal to the total thickness of the SL ($d$) leading to the linear dependence of $k_\perp$ on the number of periods.

In any case, either by looking at the minimum or the linear dependence of the $k_\perp$ with respect to $L$, to observe coherent thermal transport it is necessary that the incoming thermal wave retains its phase after it has been reflected or transmitted across the interface. This implies that the scattering mechanisms at each interface should not be purely diffusive, otherwise, the interfacial roughness or intermixing will destroy phonon coherence and phase information will be lost.[29,30] Therefore, the presence of atomically smooth interfaces becomes mandatory. However, this last point is not fully understood. Recently, numerical simulations carried out by Qiu *et al.* found the same about linear dependence of $k$ in rough periodic and aperiodic Si:Ge SLs with period thickness $L$ = 20 nm[31]. These findings were associated to the low interface densities ($1/L$ = 0.05 nm$^{-1}$) and weak disorder scattering. In this case, the dominant thermal phonons are not affected by the disorder scatterings and they can transverse ballistically the SLs regardless of aperiodicity or interface roughness. Similar results were found by Wang et al. [32] and Chakraborty et al.[33] in rough periodic SLs and random multilayer structures (RML) made of artificial atoms. Both simulations showed the same linear-like behaviour of $k_\perp$ vs $L$. However, the absence of a minimum in $k_\perp$ as a function of total thickness in the simulations performed by Wang et al. suggest a ballistic phonon transport rather than coherent effects [32].

In this work we report ultralow thermal conductivity in rough (TiNiSn):(HfNiSn) (with abbreviation (TNS):(HNS)) half-Heusler SLs. The period length of the SLs has been chosen to match crossover from incoherent to coherent transport in HH SLs.[34] The measured $k_\perp$ showed values below the amorphous limit of the effective material. As far as we know, these results are the lowest experimental values reported so far for any kind of half-Heusler (HH) compounds.

## 2. Previous results

The HH compounds investigated here are n-type narrow-band-gap semiconductors with quite large Seebeck coefficient and electrical conductivity.[35,36] However, the relatively high thermal conductivity still limits their thermoelectric performance and, hence, the industrial commercialization. For this reason, our previous studies were focused on the $k$ reduction through SL structuration. We designed three different experiments to study the impact of the period length[34,37] and the period composition[38] on the electrical and thermal properties of HH SLs. Our findings revealed a room temperature crossover from incoherent to coherent thermal transport in HH SLs. The $k_\perp$ vs $L$ exhibited a continuous diminution of $k_\perp$ as $L$ decreases, showing a minimum of $k_\perp = 1.11 \pm 0.06$ W K$^{-1}$ m$^{-1}$ at $L \sim 3.2$ nm. At smaller $L$ the $k_\perp$ rises up entering in the coherent regime.[34,37]

## 3. Experimental results and discussion

In this work, we have taken a different experimental approach to study the heat transport trough the SL. Instead of fabricating smooth and defect-free SLs, we have deteriorated the quality of the interfaces by changing the deposition conditions. We used DC magnetron sputtering to fabricate eight (TNS):(HNS) SLs with period thicknesses ranging from 2.9 nm < $L$ < 4.8 nm. The $L$ was determined from the best fit of the X-ray diffractograms using CADEM: calculate X-ray diffraction of epitaxial multilayers.[39] Five SLs were grown with the same number of periods $N = 37$ ($S_1$, $S_2$, $S_3$, $S_7$ and $S_8$, respectively). Other three samples were deposited with different number of periods $N = 111$ ($S_4$, $S_6$) and 148 ($S_5$). All these samples, except $S_1$ (homogeneous-growth), were grown 30 mm away from the center of the cathodes in the inhomogeneous part of the plasma cloud (inhomogeneous-growth). Two different deposition conditions were used here. $S_1$, $S_2$, $S_6$, $S_7$ and $S_8$ were grown at low gas pressure and cathode power (low rate), while $S_3$, $S_4$ and $S_5$ were grown at high gas pressure and cathode power (high rate). The surface roughness was determined from the root mean square of a two dimensional power spectral density plot of the sample surface measured by atomic force microscopy, AFM. For convenience, the AFM surface-roughness will be referred simply as roughness ($\eta$).**Table 1** lists a summary of all the samples measured in this work. A detailed description of the sample fabrication can be found in the supporting information.

A cross-sectional transmission electron microscope (TEM) image of one SL with a roughness of $\eta = 5.9$ nm and period thickness $L = 4.5$ nm is displayed in **Figure 1a**. As it is displayed in the inset of **Figure 1a**, there is an intermixing of the SL layers, however, the SL still keeps the crystal and epitaxial quality as shown in the rocking curve in **Figure 1b** and its inset. The rocking curves reveal the broadening of a given diffraction peak. Defects such as mosaicity, atomic intermixing dislocations, among others, lead to spreading of crystal planes and thus a broadening of the linewidth.[40] In addition, the presence of the (002) and (004) film reflections around $2\theta = 30°$ and $60°$, respectively, confirm the crystallinity of all the samples discarding amorphization of the crystal structure (see **Figure S3**, **S4**, and **S6** in the supporting information). The crystal quality can also be appreciated in the high resolution TEM image, where it is possible to observe the well-ordered crystal structure (see inset **Figure 1a**).

| Sample | Parameters | | | | | | | |
|---|---|---|---|---|---|---|---|---|
| | Ar pressure [mbar] | Cathode power [W] | | Number of periods (N) | $L$ [nm] | $\eta$ [nm] | FWHM [deg] | Thickness [nm] | Sample holder |
| | | TiNiSn | HfNiSn | | | | | | |
| $S_1$ | 0.021 | 7 | 7 | 37 | 2.9 | 0.425 | 1.31 | 108 | Single |
| $S_2$ | 0.021 | 7 | 7 | 37 | 3.5 | 0.906 | 1.53 | 107 | Single |
| $S_3$ | 0.031 | 24 | 16 | 37 | 4.5 | 5.94 | 1.08 | 159 | Single |
| $S_4$ | 0.031 | 24 | 16 | 111 | 3.9 | 27.9 | 0.77 | 450 | Single |
| $S_5$ | 0.031 | 24 | 16 | 147 | 4.8 | 28 | 0.75 | 637 | Single |
| $S_6$ | 0.021 | 7 | 7 | 111 | 4.5 | - | 1.00 | 506 | Single |
| $S_7$ | 0.021 | 7 | 7 | 37 | 4.5 | - | 1.21 | 107 | Double |
| $S_8$ | 0.021 | 7 | 7 | 37 | 4.5 | - | 1.21 | 107 | Double |

**Table 1** *Summary of the deposition parameters (Ar pressure and cathode power), number of periods, period length (L), surface roughness ($\eta$), full width half maximum of rocking curve (FWHM,) total SL thickness and sample holder of investigated samples*

The cross-plane thermal conductivity was measured using well-known three-omega ($3\omega$) method[41,42] in the differential configuration.[43,44]

First, we will focus on $S_1$ and $S_2$, grown under the same low sputtering rates but at different distance of the cathodes. The homogeneously-grown SL ($S_1$) shows significantly higher $k_\perp$ than the inhomogeneous SL ($S_2$) above 120 K. It appears that the difference in period lengths ($L_1$ = 2.9 nm and $L_2$ = 3.5 nm for $S_1$ and $S_2$, respectively) may explain this finding. However, in our previous work, we found that the $k_\perp$ decreases as period length decreases achieving a minimum value $k_\perp \approx 1.11$ W K$^{-1}$ m$^{-1}$ at $L \approx 3.2$ nm.[34] Then, as both period lengths of the SLs are located around this minimum, the $k_\perp$ should be almost identical. Therefore, the difference in $k_\perp$ cannot be associated exclusively to SL period. At first glance, the increase of $\eta$ may also explain this behavior. Moreover, after change in the deposition conditions to induce higher $\eta$ ($S_3$), the $k_\perp$ decreases even more reaching values as low as the theoretical amorphous limit of HNS and below the amorphous limit of an effective material.[45] This behavior is preserved along a wide temperature range 100 < T < 300 K as it is displayed in **Figure 2a**. A deeper description of the theoretical amorphous limit is given in section 3 of the supporting information.

As we already discarded a possible amorphization of the films and the impact of $L$, the other factor that could explain the very low $k_\perp$ is a higher concentration of defects and the loss of the epitaxial growth of $S_3$. But, the sharpness of the rocking curve of $S_3$ ($\Gamma_3 = 1.08°$) in comparison with $S_1$ ($\Gamma_1 = 1.31°$) and $S_2$ ($\Gamma_2 = 1.51°$) indicates that $S_3$ is superior in terms of epitaxial quality to the other two SLs and the $k_\perp$ of $S_3$ should be even higher. Thus, the extremely low $k_\perp$ values shown by $S_3$ can be directly related to the high increase of $\eta$. But, for the case of the other two SLs, as the FWHM of $\Gamma_1 < \Gamma_2$, it is reasonable that the difference in $k_\perp$ can be associated to a combination of both the loss of crystal quality and the increase of roughness. This dependence is better appreciated in **Figure 2b**, where $k_\perp$ is plotted as a function of $\eta$ and $\Gamma/\eta$ (bottom and top x-axis, respectively). We can see that $k_\perp$ monotonically decreases with $\eta$ from 300 K to 170 K. While for lower temperatures, we observe that $k_\perp$ is getting constant for low $\eta$ values. On the other hand, we can observe that for high roughness (or small FWHM/$\eta$) the $k_\perp$ increases dramatically. Similar behavior was also observed by Termentzidis et al. using molecular dynamics simulations. [46,47] We will return to this point later.

Now, we will analyze behavior of $k_\perp$ as the number of periods increases for the SLs grown with similar conditions as $S_3$. **Figure 3a** shows the $k_\perp$ as function of number of periods corresponding to the samples $S_3$, $S_4$ and $S_5$, respectively. The period length in this case should be equal but it is possible to see a shift of the satellite peaks (**Figure S4** supporting information). The calculated period length was found $L = 3.9$ nm and 4.8 nm for $S_4$ and $S_5$, respectively.

From **Figure 3a**, one sees that $k_\perp$ rises continuously at 250 K and at 170 K similarly to the behaviour observed by Luckyanova[30]. While for 100 K the $k_\perp$ seems to be constant for thicker samples. The nearly linear dependence $k_\perp$ on the number of periods seems to indicates that part of the heat is transported still by phonons with mean free path in the order of the sample thicknesses. Other interpretation of this phenomena can be also associated to epitaxial quality of the thicker samples. The inset of **Figure 1b** shows very sharp FWHM $S_4$ ($\Gamma_4 = 0.77°$) and $S_5$ ($\Gamma_5 = 0.75°$). This means that both $S_4$ and $S_5$ have superior epitaxial quality than $S_3$. As this effect may also play a role here, the $k_\perp$ as function of FWHM is plotted in **Figure 4a**. From this graph it is possible to observe certain correlation between $k_\perp$ and the FWHM. Except for $S_3$, there is a about-linear decrease of the $k_\perp$ as FWHM increases. Now, if we pay attention to the plot of the thermal conductance, $k_\perp \cdot d$ (where $d$ is the total thickness of each SL), vs FWHM there is a clear correlation between the samples grown under the same deposition conditions (see **Figure 4b**).

The other important parameter that we have to take into account is the surface roughness, which rises significantly with the number of periods. In contrast to previous cases, here the surface roughness of $S_4$ and $S_5$ is $\eta \sim 28$ nm, which is six times larger than the $L$ of the SL. Such huge $\eta$ should also impact on the $k_\perp$ reducing it even more, in spite of that, we observe that the experimental $k_\perp$ is still increasing (see **Figure 2b**). We can notice that the $k_\perp$ rises up for larger number of periods ($S_6$, $N = 111$ and $L = 4.5$ nm).

As we mentioned above, theoretical simulations of Termentzidis et al. [46,47] showed an increase of $k_\perp$ of SLs with very rough interfaces compared with atomically smooth SLs [46,47]. They suggested that the thermal conductivity of a SL is mainly controlled by the Kapitza resistance of interfaces, which in turn seems to be governed by the interfacial area. It is because a large majority of phonons have wavelength ($\lambda$) larger than $\eta$ and they see the interface as a planar one. Then, the transmitted heat flux is controlled by the projected area. On the other hand, for very rough interfaces, most of the phonons have much smaller $\lambda$ than the $\eta$, then, the phonons will not feel the interface as planar, the phonon scattering will be incoherent and the transmitted heat flux is controlled by real contact area of the rough interface. In other words if $\lambda > \eta$, the effective scattering area would be the projected one and $k_\perp$ decreases slightly. For $\eta \sim \lambda$ the interface will strongly scatter phonons and $k_\perp$ will decrease even further. But if $\eta > \lambda$ the interfacial scattering becomes again negligible and the transmitted energy is proportional to the true area. [47] Similar behavior is also observed in SLs grown at low deposition rate (see **Figure 3b**).

Another interesting question concerning reduction of the thermal conductivity is, whether it is caused just by the fact that the one single component of the SL grows in a special way or if it is caused by the interaction of both components. To test this effect a double substrate holder was used for deposition. On this holder, there is room for two substrates lying side by side. So, during the deposition processes, one substrate is a little bit closer to the HNS and the other to the TNS, respectively. As the samples were grown in the inhomogeneous region of the plasma cloud, one of the SLs will contain a little bit more TNS or HNS per period than the other. This effect can be appreciated in XRD diffractograms displayed in the **Figure S6** in the supporting information. The XRD diffractogram shows that the maxima of (002) HH peaks are shifted. As expected, the sample closer to HNS cathode ($S_6$) has a maximum at 29.4° which is closer to (002) peak of HNS (29.36°). While the sample closer to TNS cathode ($S_7$) has a maximum at 29.9° which is closer to (002) peak of TNS (30.06°). The rocking curves of both samples show an equal FWHM value of 1.21°. Here we focused on SLs having the same number of periods, which were grown with low sputtering rate to minimize the impact of the roughness. In **Figure 3c**, the $k_\perp$ of $S_6$ and $S_7$ are compared to the SL grown using single sample-holder $S_2$. It is clear that the sample that contained more TNS ($S_6$) has higher thermal conductivity than the sample that contained more HNS ($S_7$) and the single sample-holder ($S_2$). Similar results we also observed in our previous investigation in HHs SLs. Where we found a minimum of the $k_\perp$ ~ 1.39 W K$^{-1}$ m$^{-1}$ for SL having the same amount of each material.[38]

Coming back to the idea of coherent transport, while it is interesting to speak about a possible coherent transport in rough SLs, we cannot proof that the heat transport is influenced by coherent phonons just based on the nearly-linear dependence of $k_\perp$ on the number of periods. Other mechanism such as material intermixing, different degree of interface roughness and privileged growth of one of the constituent materials of the SL could also explain this linear dependence of $k_\perp$. Therefore, an observation of a linear increase of $k_\perp$ with the number of periods does not necessarily imply coherent transport. Rather, it can be explained by phonon mean free paths larger than the system length, i.e. ballistic transport. Furthermore, it is also important to mention the impact of low and high growth rates on the structures and their thermal conductivities. While the low rate warranties very smooth interfaces with low surface roughness, it induces a mosaic effect reducing the thermal conductivity of the samples. On the other hand the high rate reduces the mosaic effect but leads to very rough interfaces. At the very rough interfaces, the interfacial scattering becomes negligible and the transmitted energy is favoured. [47]

## 3. Conclusions

In this work we found out that the samples that were grown in the inhomogeneous region of the deposition cloud exhibited significantly lower thermal conductivity than the sample grown at the homogenous part. The thermal conductivity can be reduced even below the amorphous limit by using higher gas pressure and cathode power for deposition process. This is an outstanding result because it means that a solid body with good crystalline qualities (as implied by the quite narrow rocking curves) has a lower thermal conductivity than it should have in amorphous state. We also observed experimentally a linear-like increase of $k_\perp$ as a function of the number of periods for SL grown under variable deposition conditions. While this behavior has been reported before as coherent transport, we cannot prove that this is the case in this work. Other parameters such us the degree of intermixing, interface roughness and crystal quality may also play a role. Furthermore, we have also demonstrated that the thermal conductivity is influenced by the way in which one of the single components grows within the inhomogeneous region. Finally, our findings show a large potential for thermoelectric generators where a huge reduction of *k* is required but without losing the crystal quality of the system.

**Acknowledgements**

We gratefully acknowledge funding by the Deutsche Forschungsgemeinschaft (DFG, German Research Foundation) - project number 121583221 and project number 198551693 and by the Graduate School of Excellence Material Science in Mainz (GSC266).

**References**

[1]  M. Maldovan, "*Narrow Low-Frequency Spectrum and Heat Management by Thermocrystals*", Phys. Rev. Lett., vol. 110, no. 2, pp. 25902, 2013. DOI: 10.1103/PhysRevLett.110.025902.

[2]  S. Guenneau, C. Amra and D. Veynante, "*Transformation thermodynamics: Cloaking and concentrating heat flux.*", Opt. Express, vol. 20, no. 7, pp. 8207–18, 2012.

[3]  T. Han, X. Bai, J.T.L. Thong, B. Li and C.-W. Qiu, "*Full control and manipulation of heat signatures: cloaking, camouflage and thermal metamaterials*", Adv. Mater., vol. 26, no. 11, pp. 1731–1734, 2014. DOI: 10.1002/adma.201304448.

[4]  H. Xu, X. Shi, F. Gao, H. Sun and B. Zhang, "*Ultrathin three-dimensional thermal cloak*", Phys. Rev. Lett., vol. 112, no. 5, pp. 54301, 2014. DOI: 10.1103/PhysRevLett.112.054301.

[5]  P. Ben-Abdallah and S.-A. Biehs, "*Near-field thermal transistor*", Phys. Rev. Lett., vol. 112, no. 4, pp. 44301, 2014. DOI: 10.1103/PhysRevLett.112.044301.

[6]  B. Li, L. Wang and G. Casati, "*Thermal diode: Rectification of heat flux*", Phys. Rev. Lett., vol. 93, no. 18, pp. 184301, 2004. DOI: 10.1103/PhysRevLett.93.184301.

[7]  L. Wang and B. Li, "*Thermal memory: A storage of phononic information*", Phys. Rev. Lett., vol. 101, no. 26, pp. 267203, 2008. DOI: 10.1103/PhysRevLett.101.267203.

[8]  V. Kubytskyi, S.-A. Biehs and P. Ben-Abdallah, "*Radiative bistability and thermal memory*", Phys. Rev. Lett., vol. 113, no. 7, pp. 74301, 2014. DOI: 10.1103/PhysRevLett.113.074301.

[9]  M. Elzouka and S. Ndao, "*Near-field NanoThermoMechanical memory*", Appl. Phys. Lett., vol. 105, no. 24, pp. 243510, 2014. DOI: 10.1063/1.4904828.

[10] A.J. Minnich, "*Determining phonon mean free paths from observations of quasiballistic thermal transport*", Phys. Rev. Lett., vol. 109, no. 20, pp. 205901, 2012. DOI: 10.1103/PhysRevLett.109.205901.

[11] I. Maasilta and A.J. Minnich, "*Heat under the microscope*", Phys. Today, vol. 67, no. 8, pp. 27–32, 2014. DOI: 10.1063/PT.3.2479.

[12] G.A. Slack, *CRC Handbook of Thermoelectrics*, CRC Press, Boca Raton, 1995.

[13] K.E. Goodson, "*MATERIALS SCIENCE: Ordering Up the Minimum Thermal Conductivity of Solids*", Science, vol. 315, pp. 342–343, 2007. DOI: 10.1126/science.1138067.

[14] W.S. Capinski, H.J. Maris, T. Ruf, M. Cardona, K. Ploog and D.S. Katzer, "*Thermal-conductivity measurements of GaAs/AlAs superlattices using a picosecond optical pump-and-probe technique*", Phys. Rev. B, vol. 59, no. 12, pp. 8105–8113, 1999. DOI: 10.1103/PhysRevB.59.8105.

[15] Y.K. Koh, Y. Cao, D.G. Cahill and D. Jena, "*Heat-transport mechanisms in superlattices*", Adv. Funct. Mater., vol. 19, no. 4, pp. 610–615, 2009. DOI: 10.1002/adfm.200800984.

[16] P. Chen, N.A. Katcho, J.P. Feser, W. Li, M. Glaser, O.G. Schmidt et al., "*Role of surface-segregation-driven intermixing on the thermal transport through planar Si/Ge superlattices*", Phys. Rev. Lett., vol. 111, no. 11, pp. 115901, 2013. DOI: 10.1103/PhysRevLett.111.115901.

[17] J. Ravichandran, A.K. Yadav, R. Cheaito, P.B. Rossen, A. Soukiassian, S.J. Suresha et

al., "*Crossover from incoherent to coherent phonon scattering in epitaxial oxide superlattices*", Nat. Mater., vol. 13, no. 2, pp. 168–172, 2013. DOI: 10.1038/nmat3826.

[18] W. Kim, S. Singer, A. Majumdar, J. Zide, A. Gossard and A. Shakouri, Role of nanostructures in reducing thermal conductivity below alloy limit in crystalline solids, in ICT 2005. 24th International Conference on Thermoelectrics, 2005, pp. 9–12.

[19] R.M. Costescu, D.G. Cahill, F.H. Fabreguette, Z.A. Sechrist and S.M. George, "*Ultra-Low Thermal Conductivity in W/Al2O3 Nanolaminates*", Science, vol. 303, no. 5660, pp. 989–990, 2004. DOI: 10.1126/science.1093711.

[20] G. Pernot, M. Stoffel, I. Savic, F. Pezzoli, P. Chen, G. Savelli et al., "*Precise control of thermal conductivity at the nanoscale through individual phonon-scattering barriers*", Nat. Mater., vol. 9, no. 6, pp. 491–495, 2010.

[21] J.-P. Niemelä, A. Giri, P.E. Hopkins and M. Karppinen, "*Ultra-low thermal conductivity in TiO 2 :C superlattices*", J. Mater. Chem. A, vol. 3, no. 21, pp. 11527–11532, 2015. DOI: 10.1039/C5TA01719J.

[22] C. Chiritescu, D.G. Cahill, N. Nguyen, D. Johnson, A. Bodapati, P. Keblinski et al., "*Ultralow thermal conductivity in disordered, layered WSe2 crystals.*", Science, vol. 315, no. 5810, pp. 351–3, 2007. DOI: 10.1126/science.1136494.

[23] J. Garg and G. Chen, "*Minimum thermal conductivity in superlattices: A first-principles formalism*", Phys. Rev. B, vol. 87, no. 14, pp. 140302, 2013. DOI: 10.1103/PhysRevB.87.140302.

[24] Y. Wang, H. Huang and X. Ruan, "*Decomposition of coherent and incoherent phonon conduction in superlattices and random multilayers*", Phys. Rev. B, vol. 90, no. 16, pp. 165406, 2014. DOI: 10.1103/PhysRevB.90.165406.

[25] H. Mizuno, S. Mossa and J.-L. Barrat, "*Beating the amorphous limit in thermal conductivity by superlattices design*", Sci. Rep., vol. 5, pp. 14116, 2015. DOI: 10.1038/srep14116.

DOI: 10.1080/15567265.2018.1505987[26]   R. Venkatasubramanian, "*Lattice thermal conductivity reduction and phonon localizationlike behavior in superlattice structures*", Phys. Rev. B, vol. 61, no. 4, pp. 3091–3097, 2000.

[27]   B. Saha, Y.R. Koh, J. Comparan, S. Sadasivam, J.L. Schroeder, M. Garbrecht et al., "*Cross-plane thermal conductivity of (Ti,W)N/(Al,Sc)N metal/semiconductor superlattices*", Phys. Rev. B, vol. 93, no. 4, pp. 45311, 2016. DOI: 10.1103/PhysRevB.93.045311.

[28]   M.N. Luckyanova, J. Garg, K. Esfarjani, A. Jandl, M.T. Bulsara, A.J. Schmidt et al., "*Coherent phonon heat conduction in superlattices*", Science, vol. 338, no. 6109, pp. 936–939, 2012. DOI: 10.1126/science.1225549.

[29]   M. Maldovan, "*Phonon wave interference and thermal bandgap materials*", Nat. Mater., vol. 14, no. 7, pp. 667–674, 2015. DOI: 10.1038/nmat4308.

[30]   S. Merabia and K. Termentzidis, *Nanostructures and Heat Transport*, in *Nanostructured Semiconductors: Amorphization and Thermal Properties*, K. Termentzidis, ed., Pan Stanford Publishing Pte. Ltd., 2017, pp. 69–99.

[31]   B. Qiu, G. Chen and Z. Tian, "*Effects of aperiodicity and roughness on coherent heat conduction in superlattices*", Nanoscale Microscale Thermophys. Eng., vol. 19, no. 4, pp. 272–278, 2015. DOI: 10.1080/15567265.2015.1102186.

[32]   Y. Wang, C. Gu and X. Ruan, "*Optimization of the random multilayer structure to break the random-alloy limit of thermal conductivity*", Appl. Phys. Lett., vol. 106, no. 7, pp. 73104, 2015. DOI: 10.1063/1.4913319.

[33]   P. Chakraborty, L. Cao and Y. Wang, "*Ultralow Lattice Thermal Conductivity of the Random Multilayer Structure with Lattice Imperfections*", Sci. Rep., vol. 7, no. 1, pp. 8134, 2017. DOI: 10.1038/s41598-017-08359-2.

[34]   P. Hołuj, C. Euler, B. Balke, U. Kolb, G. Fiedler, M.M. Müller et al., "*Reduced thermal conductivity of TiNiSn/HfNiSn superlattices*", Phys. Rev. B, vol. 92, no. 12, pp. 125436,
14

2015. DOI: 10.1103/PhysRevB.92.125436.

[35] G.J. Poon, *Electronic and thermoelectric properties of Half-Heusler alloys*, in *Semiconductors and semimetals: Recent trends in thermoelectric materials research II Vol. 70*, T.M. Tritt, ed., Academic Press, 2001, pp. 37–75.

[36] R.A. Downie, S.A. Barczak, R.I. Smith and J.W.G. Bos, "*Compositions and thermoelectric properties of XNiSn (X = Ti, Zr, Hf) half-Heusler alloys*", J. Mater. Chem. C, vol. 3, no. 40, pp. 10534–10542, 2015. DOI: 10.1039/C5TC02025E.

[37] P. Komar, T. Jaeger, C. Euler, E. Chávez Angel, U. Kolb, M.M. Müller et al., "*Half-Heusler superlattices as model systems for nanostructured thermoelectrics*", Phys. status solidi, vol. 213, pp. 732–738, 2015. DOI: 10.1002/pssa.201532445.

[38] P. Komar, E. Chávez-Ángel, C. Euler, B. Balke, U. Kolb, M.M.. Müller et al., "*Tailoring of the electrical and thermal properties using ultra-short period non-symmetric superlattices*", APL Mater., vol. 4, no. 10, pp. 104902, 2016. DOI: 10.1063/1.4954499.

[39] P. Komar and G. Jakob, "*CADEM : calculate X-ray diffraction of epitaxial multilayers*", J. Appl. Crystallogr., vol. 50, no. 1, pp. 288–292, 2017. DOI: 10.1107/S1600576716018379.

[40] C.G. Darwin, "*XCII. The reflexion of X-rays from imperfect crystals*", London, Edinburgh, Dublin Philos. Mag. J. Sci., vol. 43, no. 257, pp. 800–829, 1922. DOI: 10.1080/14786442208633940.

[41] D.G. Cahill, "*Thermal conductivity measurement from 30 to 750 K: the 3ω method*", Rev. Sci. Instrum., vol. 61, no. 2, pp. 802, 1990. DOI: 10.1063/1.1141498.

[42] D.G. Cahill, "*Erratum: "Thermal conductivity measurement from 30 to 750 K: The 3ω method" [Rev. Sci. Instrum. 61, 802 (1990)]*", Rev. Sci. Instrum., vol. 73, no. 10, pp. 3701, 2002. DOI: 10.1063/1.1505652.

[43] D. Cahill, M. Katiyar and J. Abelson, "*Thermal conductivity of a-Si:H thin films*", Phys.

DOI: 10.1080/15567265.2018.1505987Rev. B, vol. 50, no. 9, pp. 6077–6081, 1994. DOI: 10.1103/PhysRevB.50.6077.

[44] J.H. Kim, A. Feldman and D. Novotny, "*Application of the three omega thermal conductivity measurement method to a film on a substrate of finite thickness*", J. Appl. Phys., vol. 86, no. 7, pp. 3959–3963, 1999. DOI: 10.1063/1.371314.

[45] D.G. Cahill, S.K. Watson and R.O. Pohl, "*Lower limit to the thermal conductivity of disordered crystals*", Phys. Rev. B, vol. 46, no. 10, pp. 6131–6140, 1992. DOI: 10.1103/PhysRevB.46.6131.

[46] K. Termentzidis, S. Merabia, P. Chantrenne and P. Keblinski, "*Cross-plane thermal conductivity of superlattices with rough interfaces using equilibrium and non-equilibrium molecular dynamics*", Int. J. Heat Mass Transf., vol. 54, no. 9–10, pp. 2014–2020, 2011. DOI: 10.1016/j.ijheatmasstransfer.2011.01.001.

[47] S. Merabia and K. Termentzidis, "*Thermal boundary conductance across rough interfaces probed by molecular dynamics*", Phys. Rev. B, vol. 89, no. 5, pp. 54309, 2014. DOI: 10.1103/PhysRevB.89.054309.
**Figures**

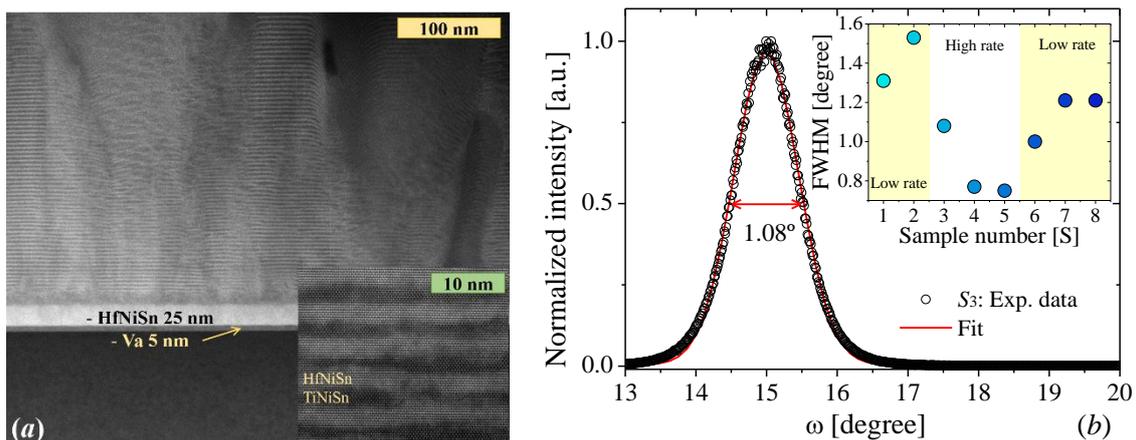

**Figure 1.** (*a*) Cross-sectional TEM image of 37 periods and L ~ 4.5 nm SL. **(inset)** High-Resolution TEM (HRTEM) image of one part of the SL. (*b*) Measured rocking curve of HfNiSn (002) peak of S3 SL. **(inset)** Full-width-half-maximum (FWHM or $\Gamma$) of the rocking curve of each sample investigated in this work.

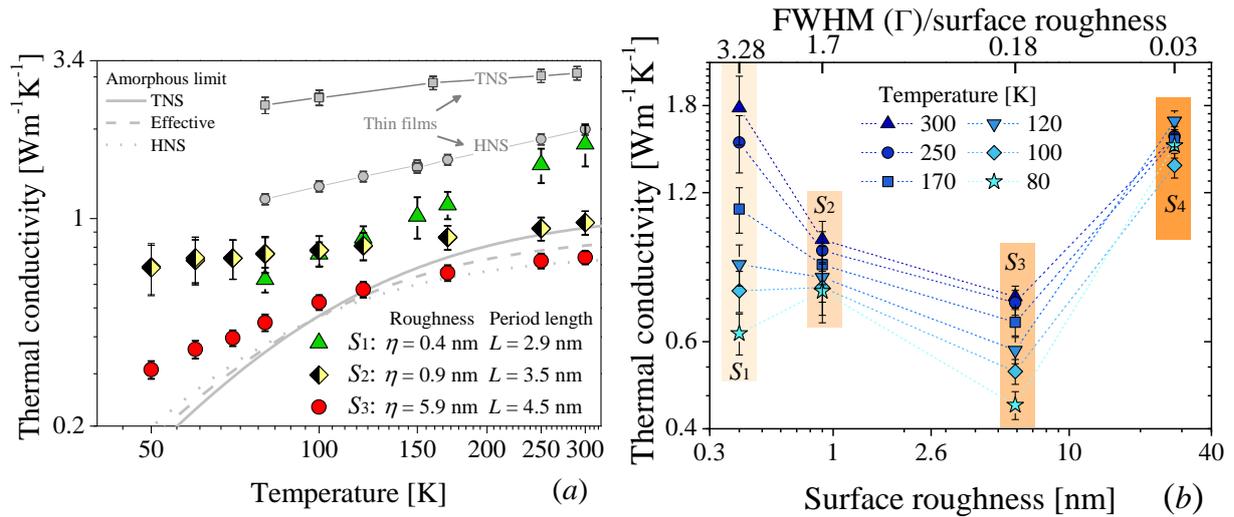

**Figure 2 (a)** Temperature dependence of the $k_⊥$ of three different SLs (colored solid symbols) and of two 1000 nm thick TNS and HNS thin films (grey square and circle symbols, respectively). The solid and dotted grey lines represent the theoretical amorphous limit of TNS and HNS. The dashed grey line represents the amorphous limit of an effective material composed by mixture of both HHs. **(b)** Cross-plane thermal conductivity as a function of the ratio of FWHM of the rocking curve and surface roughness measured at six different temperatures.

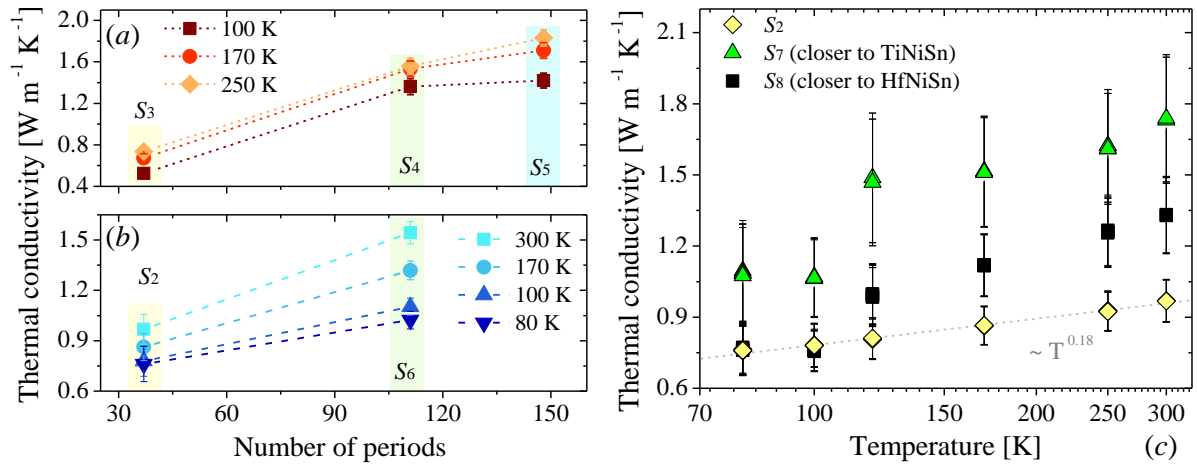

**Figure 3 (a-b)** Thermal conductivity vs number of periods measured at different temperatures for: SLs fabricated using **(a)** high and **(b)** low deposition rates, respectively. **(c)** Temperature dependence of the thermal conductivity for three different SLs grown at low deposition rates and using different sample-holders.

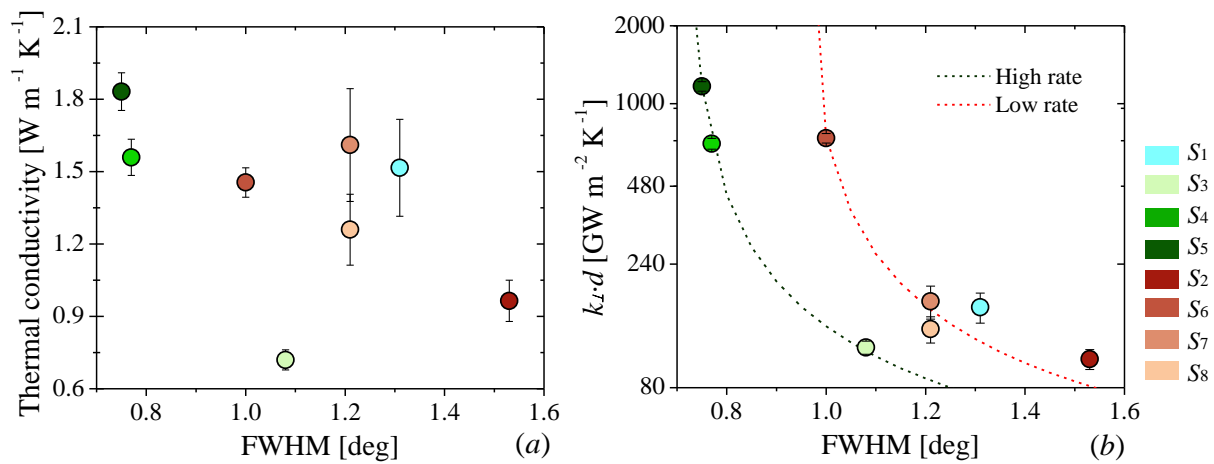

**Figure 4** (*a*) Cross-plane thermal conductivity and (*b*) thermal conductance ($k_\perp \cdot d$) as a function of FWHM of the rocking curves measured at T = 250 K. The dashed dark-green and red lines are used to guide the eye.

# Supporting Information

## Subamorphous thermal conductivity of crystalline half-Heusler superlattices

E. Chavez-Angel[1,*,†], N. Reuter[1*], P. Komar[1,2,‡], S. Heinz[1,2], U. Kolb[3,4], H.-J. Kleebe[4] and G. Jakob[1,2]

[1] Institut für Physik, Johannes Gutenberg Universität Mainz, Staudingerweg 7, 55128 Mainz, Germany.
[2] Graduate School Materials Science in Mainz, Staudingerweg 9, 55128 Mainz, Germany
[3] Institute of Inorganic and Analytical Chemistry, Johannes Gutenberg Universität Mainz, Duesbergweg 10-14, 55128 Mainz, Germany.
[4] Institute of Applied Geosciences, TU Darmstadt, Schnittspahnstraße 9, 64287 Darmstadt, Germany.

E-mail corresponding author: emigdio.chavez@icn2.cat

Keywords: Ultralow thermal conductivity, superlattices, amorphous limit of thermal conductivity

## Nomenclature

| | | | | |
|---|---|---|---|---|
| $a$ | Lattice constant [m] | | *Greek symbols* | |
| AC | Alternating current | | $\delta$ | Error deviation |
| AFM | Atomic force microscopy | | $\phi$ | Phase lag |
| $b$ | Half width of three-omega heater [m] | | $\eta$ | Surface roughness [nm] |
| $d$ | Thickness of sample or distance to the centre of the cathode [m] | | $\Lambda$ | Phonon mean free path [m] |
| DC | Direct current | | $\lambda$ | Thermal wavelength and/or wavelength of the heat carrier [m] |
| $f$ | Frequency [Hz] | | $\Theta_D$ | Debye temperature [K] |
| $h$ | Planck constant [J s] | | $\omega$ | Angular frequency ($2\pi f$) [rad s$^{-1}$] |
| $i$ | Imaginary number | | | |
| $I$ | Current [A] | | *Subscripts* | |
| $k$ | Thermal conductivity [W K$^{-1}$ m$^{-1}$] | | $app$ | Applied current [A] |
| $k_B$ | Boltzmann constant [J K$^{-1}$] | | $c$ | Corrected temperature |
| $l$ | Heater length [m] | | $iso$ | Insolation layer |
| MFP | Mean free path | | $f$ | Thin film |
| $N$ | Number of periods | | $h$ | heater |
| $n$ | Number density of atoms [m$^{-3}$] | | $L$ | Longitudinal polarization |
| P | Power [W] | | $r$ | Reference film |
| p | Pressure [mbar] | | $rel$ | Relative error deviation |
| $R$ | Resistance [Ω] | | $rms$ | Root mean square |
| $SL$ | Superlattice | | $S$ | Substrate |
| $T$ | Temperature [K] | | $SLs$ | Superlattices |
| $U$ | Voltage [V] | | $sys$ | Full system including insulation layer, film of interest, reference and the substrate |
| $v$ | Sound velocity [m s$^{-1}$] | | $T$ | Transverse polarization |
| XRR | X-ray reflectivity | | | |

---

[*] Equally contributing authors
**Current address**
[†] Catalan Institute of Nanoscience and Nanotechnology (ICN2), CSIC and BIST, Campus UAB, Bellaterra, 08193 Barcelona, Spain.
[‡] Photonics Group, Institute of Physics, Lodz University of Technology, Wólczańska 219, 90-924 Łódź, Poland.

# Supporting Information

**Characterization techniques**

For structural characterization we employed Cs corrected scanning transmission electron microscope (STEM), X-ray diffraction (XRD) and atomic force microscopy (AFM). The high resolution STEM (HR-STEM) measurements were performed on tripod polished samples using JEOL JEM ARM 200F operated at 200 kV, applying high angle annular dark field (HAADF) imaging. The crystallographic quality was determined based on $\theta-2\theta$ patterns and $\omega$ (rocking curve) scans recorded by a Bruker D8 Discovery X-ray diffractometer operated in Bragg-Brentano geometry. Surface roughness was measured by the root mean square (RMS) of a two dimensional power spectral density plot in a representative range of the sample's surface. It was recorded by Veeco Dimension 3100 setup, operated in the contact mode. The samples thickness were measured by using the same AFM equipment.

1. **Fabrication process**

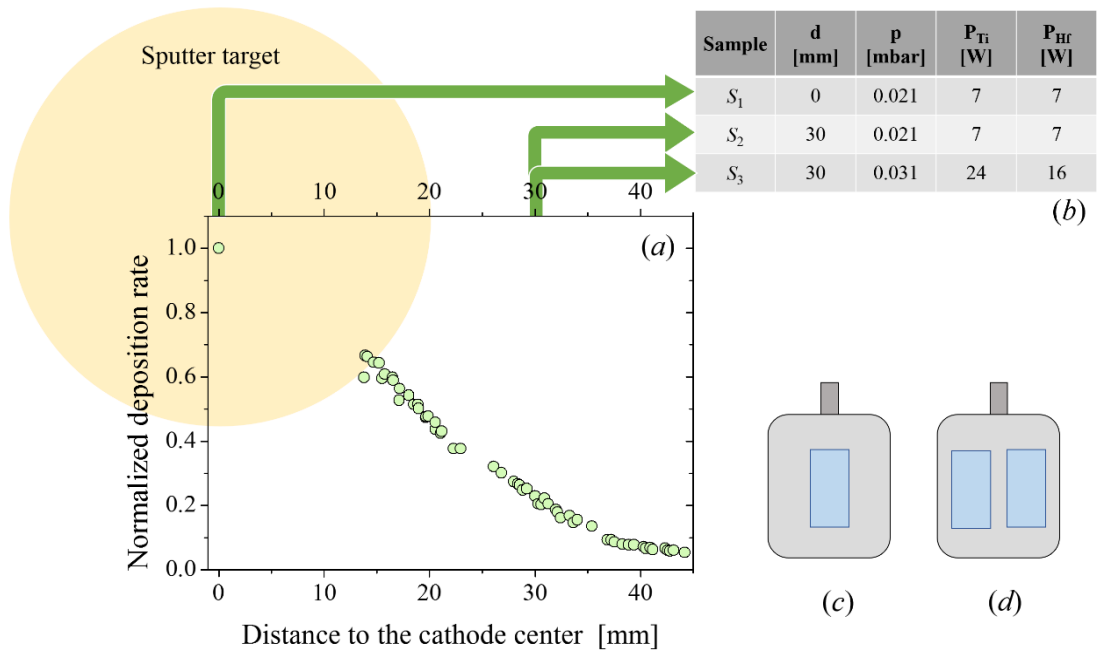

**Figure S1** (*a*) *Normalized deposition rate of the used TiNiSn and HfNiSn cathodes as a function of distance between cathode center line and MgO substrate center* (*b*) *The creation parameters for each of the three samples.* (*c*)-(*d*) *Sample holder for a single* (*c*) *and double* (*d*) *SL deposition.*

The SLs were grown by DC magnetron sputtering processes on 10 mm×5 mm MgO substrates using a Vanadium (5 nm) and HfNiSn (25 nm) as a buffer layer at T = 520 °C. The deposition rate was measured using AFM and XRR. As it is displayed in Figure S1, the deposition rate depends on the distance between substrate and central axis of the cathode due to the spiral trajectories that the ionized gas atoms follow around the field lines of the inhomogeneous magnetic field during the sputtering process. The SLs were grown at two different positions in relation to the centre of the cathode using different gas pressure, cathode powers and sample holders.

# Supporting Information

The first sample ($S_1$) was grown above the cathode centre at low Ar pressure (p = 0.21 mbar), low cathode powers ($P_{Ti} = P_{Hf} = 7$ W) and using a single sample holder (see **Figure S1 *c***). The second sample ($S_2$) was deposited using the same growing conditions but 30 mm away from the centre of the cathode. The third sample ($S_3$) was grown at the same place of $S_2$ but with higher Ar pressure (p = 0.21 mbar) and cathode power ($P_{Ti} = 24$ W and $P_{Hf} = 16$ W). For these three samples ($S_1$-$S_3$), we kept constant the total number of periods $N = 37$. Other two samples were also grown using the same deposition condition than $S_3$ but with different number of periods $N = 111$ ($S_4$) and 148 ($S_5$), respectively. In addition, other two samples were grown using the same deposition condition of $S_3$ but with a different sample holder allowing to place two substrates at the same time (see **Figure S1 *d***). As one can see in the **Figure S2**, if we use the double sample holder it is clear to see that one of the substrate we will a little bit closer to one of the cathode. The thickness of all the samples presented in this work was measured by using AFM. A summary of the growth conditions is displayed in the **Table S1**.

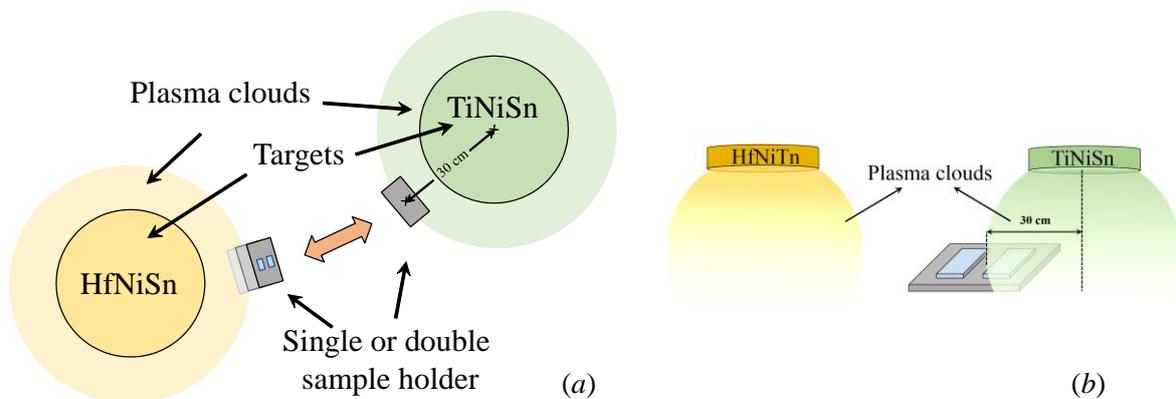

**Figure S2 (*a*)** and **(*b*)** *schematic view of the sample deposition and the growth in the inhomogeneous part of the plasma cloud.*

## 1.1 Surface and crystal structure analysis

By looking at the AFM surface scans, displayed in **Figure S3 *g*** to ***i***, one can notice a remarkable difference between the samples at a microscopic scale. Clearly, the samples grown 30 mm away from the cathode are rougher than the ones grown when the sample holder was centred. Additionally, the gas pressure and the power applied to the cathodes have also an impact on the increase of the surface roughness of the samples, as it is shown in the **Figure S3 *i***.

Regarding the XRD analysis, $S_1$ shows satellite peaks that are characteristic for SLs as shown in **Figure S3 *a***. The satellite peaks of $S_2$ and $S_3$ are less pronounced and less symmetric as expected for samples that ware sputtered at a place with spatially inhomogeneous deposition rates leading to unclear interfaces (see **Figure S3 *b*** and ***c***, respectively). Consequently, the samples are definitely different in the SL structure. However, the crystal quality of all samples is similar as one can see in the Full-Width-Half-Maximum (FWHM) of rocking curves (**Figure S3 *d*** to ***f***).

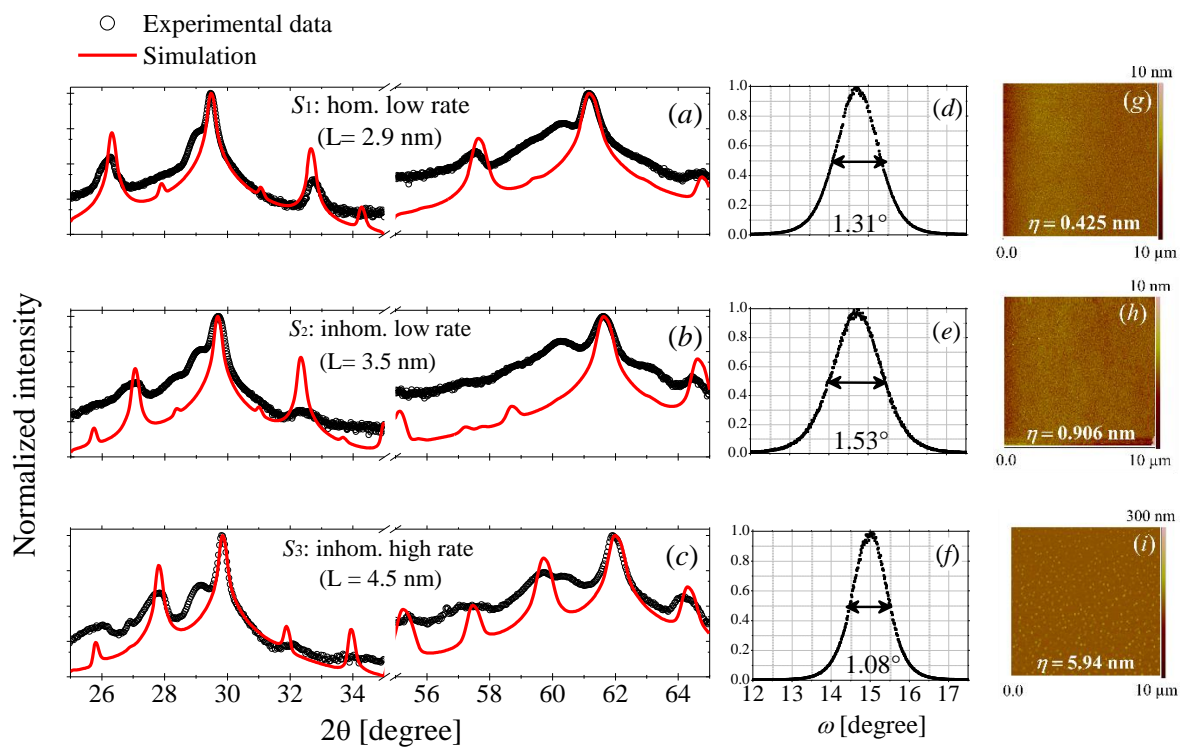

**Figure S3** *XRD diffractograms (**a**) to (**c**), rocking curves (**d**) to (**f**) and AFM surface scans (**g**) to (**h**) of 37 periods half-Heusler SLs grown under different deposition conditions. The calculation of the XRD spectra (red solid lines) was obtained by using CADEM: calculate X-ray diffraction of epitaxial multilayers.*[1]

Additionally, the XRD diffractograms of SLs with 111 and 148 periods ($S_4$ and $S_5$, respectively) are displayed **Figure S4**. These SLs were grown 30 cm away from the cathode centres with the deposition conditions identical to ones used for $S_3$.

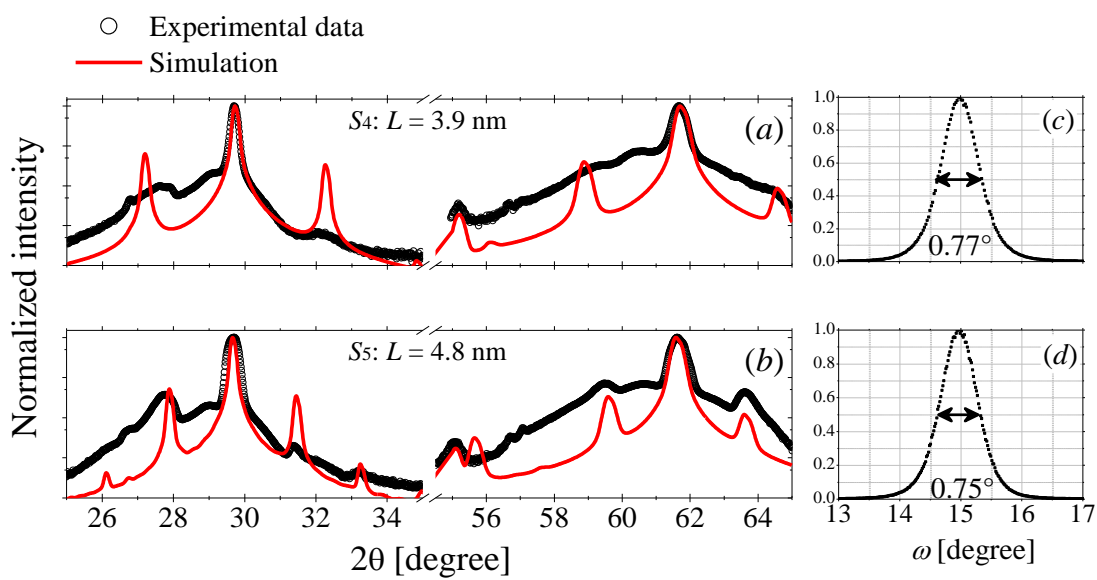

**Figure S4** *XRD diffractograms (**a**) and (**b**), rocking curves (**c**) and (**d**) of 111 ($S_4$: **a** and **c**) and 147 ($S_5$: **b** and **d**) periods half-Heusler SLs grown under the same deposition conditions as for $S_3$. The best fitting theoretical diffractogram models are shown with red lines.*[1]

# Supporting Information

As one can see in **Figure S4** the satellite peaks are weakly accented and the FWHM of the rocking curves decreases with increasing number of periods. That means that the epitaxial quality of the upper parts of the samples must be higher than the one of the lower parts. The period length in this case should be similar but one sees still shifts of the satellite peaks. The roughness rises significantly with the number of periods as one sees in the **Figure S5**.

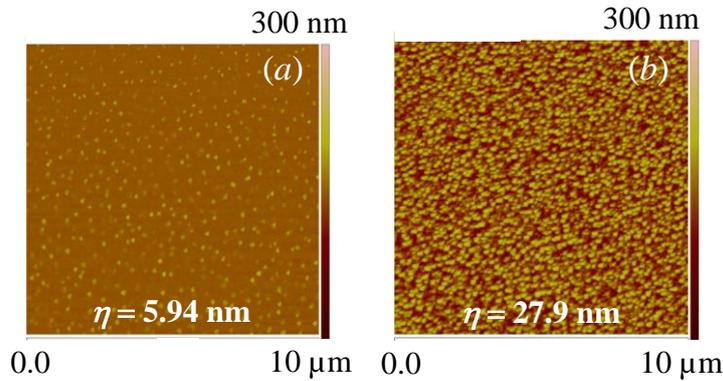

**Figure S5** *AFM pictures of superlattices grown at high rate parameters in the inhomogeneous region containing: (**a**) 37, $S_3$, and (**b**) 111, $S_4$, periods, respectively.*

Finally, the last set of samples was grown using the same conditions as for $S_2$ but with larger number of periods $N = 111$ ($S_6$) in single sample holder and using a double sample holder ($S_7$ and $S_8$) keeping $N = 37$. In the double sample holder, there is room for two substrates lying side by side so that one substrate was closer to the HfNiSn cathode and the other to the TiNiSn cathode during deposition process. As the samples were grown in the inhomogeneous region, one of the SLs will contain more TiNiSn per period and the other will contain more HfNiSn per period.

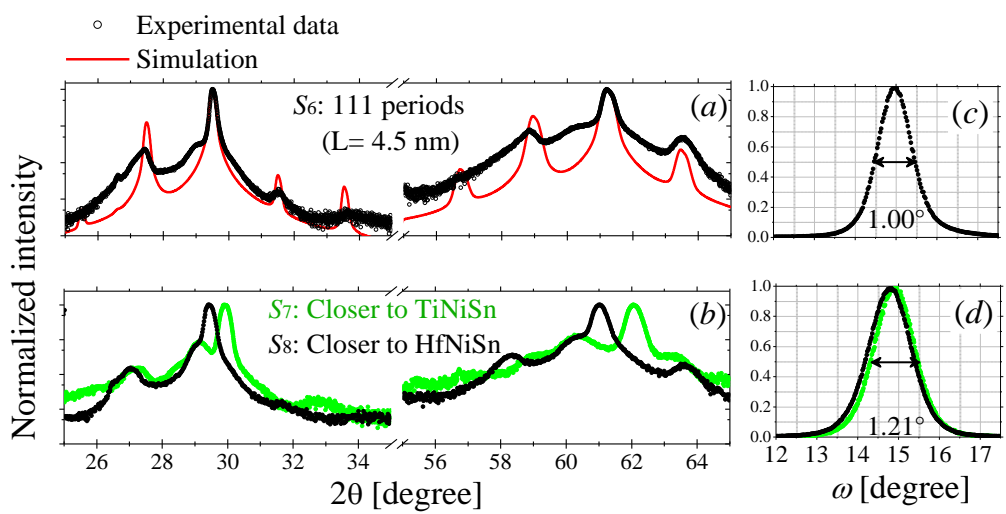

**Figure S6** *XRD Diffractograms (**a**) and (**b**) and rocking curves (**c** to **d**) of 111 ($S_6$: **a** and **c**) and 37 ($S_5$: **b** and **d**) periods half-Heusler SLs grown under the same deposition conditions of $S_2$. The best fitting theoretical diffractogram models are shown with red lines.*[1]

# Supporting Information

The period length of the samples measured here was determined from the best fit of the XRD using CADEM: calculate X-ray diffraction of epitaxial multilayers.[1] Open source and code software to calculate XRD diffractogram of any arbitrary multilayer structure. A summary of all the samples measured in this work is given in **Table S1**.

**Table S1** *Summary of the deposition parameters, surface roughness, FWHM, and total thickness of investigated samples.*

| Parameter \ Sample | d [mm] | p [mbar] | P [W] TNS | P [W] HNS | N | L [nm] | η [nm] | FWHM [°] | Thickness [nm] | Sample holder |
|---|---|---|---|---|---|---|---|---|---|---|
| $S_1$ | 0 | 0.021 | 7 | 7 | 37 | 2.9 | 0.425 | 1.31 | 108 | Single |
| $S_2$ | 30 | 0.021 | 7 | 7 | 37 | 3.5 | 0.906 | 1.53 | 107 | Single |
| $S_3$ | 30 | 0.031 | 24 | 16 | 37 | 4.5 | 5.94 | 1.08 | 159 | Single |
| $S_4$ | 30 | 0.031 | 24 | 16 | 111 | 3.9 | 27.9 | 0.77 | 450 | Single |
| $S_5$ | 30 | 0.031 | 24 | 16 | 147 | 4.8 | 28 | 0.75 | 637 | Single |
| $S_6$ | 30 | 0.021 | 7 | 7 | 111 | 4.5 | - | 1.00 | 506 | Single |
| $S_7$ | 30 | 0.021 | 7 | 7 | 37 | 4.5 | - | 1.21 | 107 | Double |
| $S_8$ | 30 | 0.021 | 7 | 7 | 37 | 4.5 | - | 1.21 | 107 | Double |

## 2. Three-omega method

The three-omega ($3\omega$) method is an electrothermal technique widely used to determine the thermal conductivity of a specimen. The experiments are performed by inducing harmonic Joule heating in a narrow metal line ($3\omega$-heater), deposited onto the surface of the sample. The metallic strip acts simultaneously as a heater and thermometer due to its temperature dependent electrical resistance as it is displayed in **Figure S7 *d***.

In our case, the $3\omega$-heater was patterned by photolithography and etching of a 50 nm thick gold thin film, grown in situ just after the deposition of $AlO_x$ insulation layer. A schematic representation and a real picture of one the samples is displayed in **Figure S7 *a*** and ***b***, respectively. The deposited metallic strip is composed of four rectangular pads connected by pins to the narrow heating wire. The width of the heating line is defined as $2b = 20$ μm and the length as $l = 1$ mm, the latter being determined by the distance between the inner pads. The outer two pads are used to apply the AC electrical current that generates the Joule heating. The inner two pads are used to measure the voltage, which contains the third harmonic component. In the experiments, a sinusoidal electrical current is applied through the resistive strip as:

$$I_{app}(t) = I_0 \cos(\omega t) \qquad (1)$$

where $I_0$ is the amplitude of the signal.

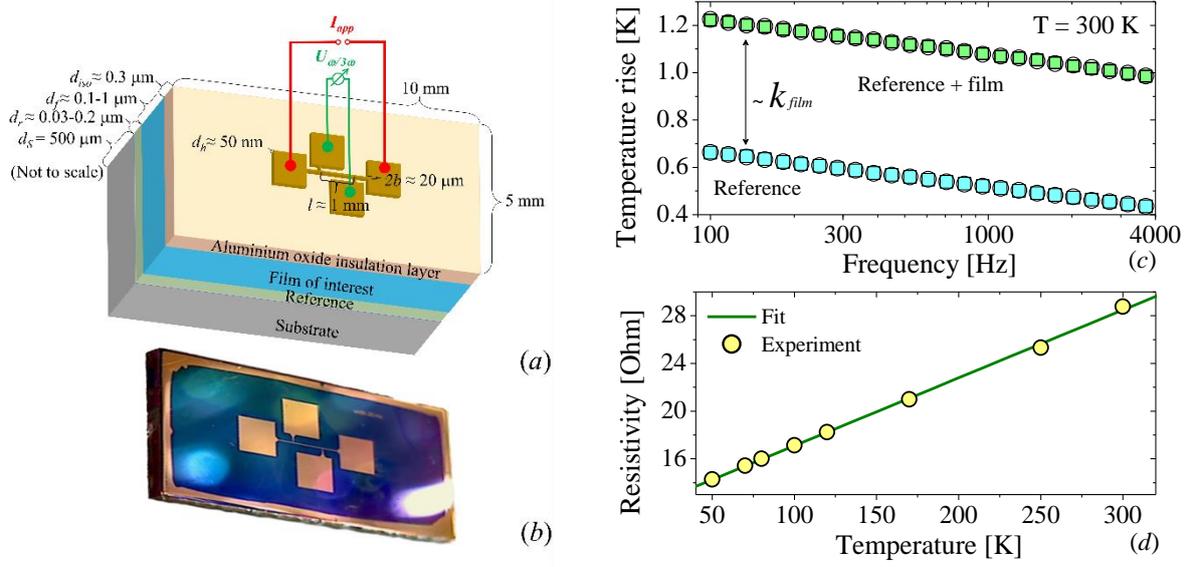

**Figure S7** (***a***) *Schematic representation of the samples measured in this work and* (***b***) *real picture of one of them.* (***c***) *Temperature rise of 3ω-heater vs frequency of 1000 μm thick HfNiSn thin film (green squares and empty dots) and its reference (blue square and empty dots).* (***d***) *Typical temperature dependence of the electrical resistivity of a 3ω-heater.*

By Joule effect, this excitation results in power dissipation that consists of a DC and AC components given by:

$$P(t) = I_{app}(t)^2 R = \frac{I_0^2 R}{2}\left(1 + \cos(2\omega t)\right) \qquad (2)$$

where $R_0$ is the resistance of the strip. As the dissipated power has a DC and AC component, the heat dissipation will result in a temperature rise that has a DC ($\Delta T_{DC}$) and an AC ($\Delta T_{AC}$) component. The temperature fluctuation of amplitude $\Delta T_{2\omega}$ will also oscillate at the same frequency.

$$T(t) = \Delta T_{DC} + \Delta T_{AC}\cos(2\omega t + \phi) \qquad (3)$$

where $\phi$ is the phase lag. Since the electrical resistivity is linearly proportional to the temperature (see **Figure S7** *c*), the $\Delta T$ will also produce a $2\omega$ oscillation in the resistivity as:

$$R(T,t) \approx R_0\left(1 + \beta\Delta T_{DC} + \beta\Delta T_{AC}\cos(2\omega t + \phi)\right) \qquad (4)$$

where $\beta$ is the temperature coefficient of the electrical resistivity of the strip. Now, by applying the Ohm's law, we obtain the modulation of the voltage of the form:

$$\begin{aligned} U &= R(T,t)I(t) \\ &= I_0 R_0 \cos(\omega t)\left(1 + \beta \Delta T_{DC} + \beta \Delta T_{AC} \cos(2\omega t + \phi)\right) \\ &= U_0 (1 + \beta \Delta T_{DC})\cos(\omega t) + \frac{U_0 \beta \Delta T_{AC}}{2}\left(\cos(\omega t + \phi) + \cos(3\omega t + \phi)\right) \end{aligned} \quad (5)$$

From (5), one is able to infer the temperature oscillations by measuring the voltage signal at the 3ω frequency [2,3]:

$$\Delta T_{AC} = \Delta T_{2\omega} = \frac{2U_{3\omega}}{\beta U_0} \approx \frac{2U_{3\omega,rms}}{\beta U_{\omega,rms}} \quad (6)$$

Since the $3\omega$ response of the voltage is very small in comparison with $1\omega$, the lock-in technique is required to extract the signal. The thermal fluctuation can therefore be obtained from the $3\omega$ component in terms of root mean square quantities (rms), as usually measured by lock-in amplifier. Due to the difference among $1\omega$ and $3\omega$ is several orders of magnitude, the noise of the whole $1\omega$ signal is in the same order as the $3\omega$ signal itself. To avoid this problem, $U_{3\omega}$ is not measured directly at the inner pads of the heater but rather with a passive circuit.

The thermal conductivity can be obtained by solving the transient heat conduction equation for a finite width line heater, deposited onto semi-infinite surface of a film-on substrate system. The temperature rise is given by:

$$\Delta T_{2\omega} = \frac{P}{lk\pi}\int_0^\infty \frac{\sin^2(xb)}{(xb)^2\sqrt{x^2+q^2}}dx \quad (7)$$

where $P$ is the applied power, $q \equiv 1/\lambda = \sqrt{2\omega/\alpha}$ is the inverse of the thermal penetration depth ($\lambda$), $\alpha$ is the thermal diffusivity and $k$ is the thermal conductivity of the material. The Eq. (7) does not have an analytical solution, however, Cahill[2,3] showed that for $\lambda \gg b$ the heater can be seen as line source. Then, the upper limit of the integral can be replaced by $1/b$ and the sinusoidal term goes $\sin(xb)/(xb) \sim 1$ in the limit of $b \to 0$. By introducing these approximations, the analytical solution is given by:

$$\Delta T_{2\omega} = \frac{P}{2lk\pi}\left(-\ln(2\omega) + \ln\left(\frac{k}{\alpha b^2}\right) + 2\gamma\right) - \frac{iP}{4kl} \quad (8)$$

where $\gamma$ is constant. Finally, the $k$ can be extracted from the slope of the real part of temperature rise vs $\ln(2\omega)$:

$$k \approx \frac{P}{2\pi l}\left(\frac{d(\Delta T_{2\omega})}{d\ln(2\omega)}\right)^{-1} \quad (9)$$

This approximation of the $3\omega$ measurement is known as slope method. In the following section the errors associated to the slope method are discussed and analyzed for our particular case. For an extended and detailed description on the derivation errors, mathematical expressions and the methodology used to calculate it, the readers are referred to the work of H.S. Carslaw and J.C. Jaeger[4], D. Cahill [2,3], Borca-Tasciuc et al.[5] C. Dames[6] and references therein.

## 2.1 Errors from mathematical description

There are three main requirements that the system has to hold to apply directly the slope method, those are: the heater is a line source, the substrate thickness is semi-infinite and the heater is infinitely long. As the real heater is not infinitesimal narrow and infinitely long in comparison to finite thick substrate. There are some limits where these considerations are valid and they are summarized in the Figure S8. As one can sees in Figure S8, in our measurements, it is always possible to choose frequencies for the $3\omega$ method to fulfil the criteria needed for the slope method, with errors below 5%.

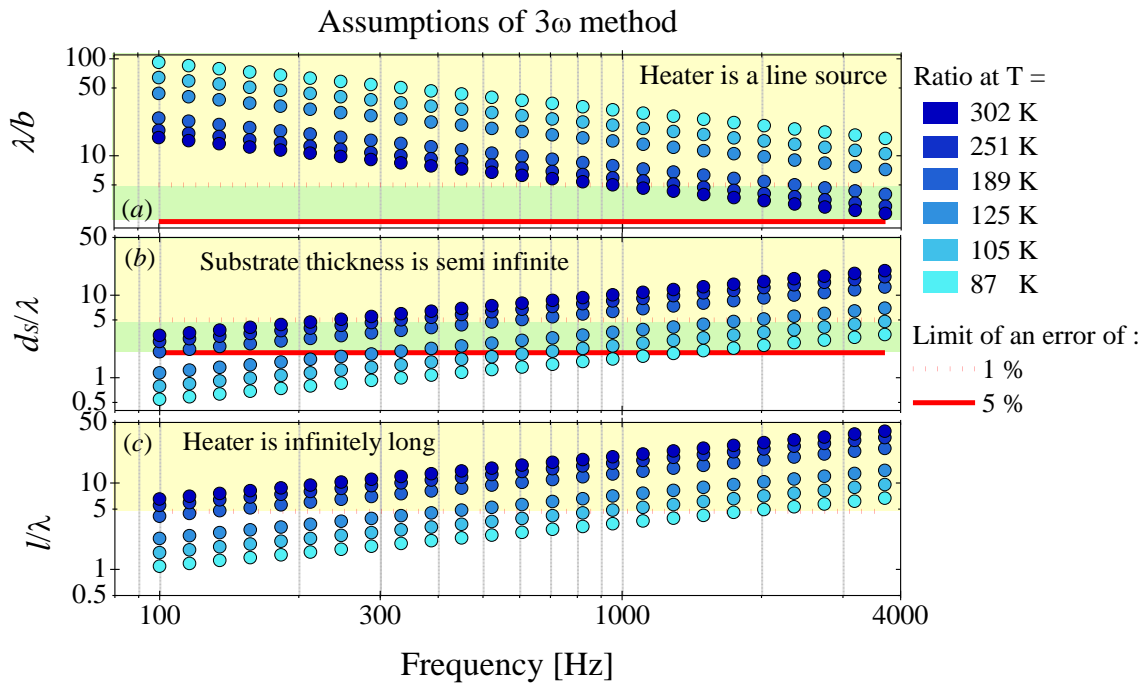

**Figure S8** *Calculated ratios and criteria for the adequacy of the applied mathematical model for our $3\omega$ measurements: (**a**) the heater is a line source, (**b**) the substrate thickness is infinite and (**c**) the heater line is infinitely long. A point fulfils a criteria if it is above the red lines in the yellow or green rectangle for an error below 1% and 5%, respectively. The substrate capacity and conductivity values that were needed to calculate the substrate penetration depth were taken from reference.[7] The criteria are taken from Ref.[6].*

## 2.2 Differential method: determination of the *k* of a thin film

Once we ensure that the slope method can be applied in our substrates, the next step is calculation estimation of $k$ of the film of interest by using the $3\omega$ differential method.[8,9] In case of a film that has a conductivity much smaller than the substrate and a heater width that is larger than the film thickness, one can model the film as a frequency independent resistance where the bigger part of the heat flows cross plane from heater-film-interface to film-substrate-interface.[8] In this case the Fourier law can be applied in one dimension:

$$P = Q = kA_{int}\frac{\Delta T_f}{d_f} \approx 2klb\frac{\Delta T_f}{d_f} \tag{10}$$

$$\Rightarrow k = \frac{Pd_f}{2lb\Delta T_f}$$

where $Q$ is the modulus of the heat flow, $A_{int}$ is the area below the heater strip ($A_{int} = 2lb$) and $\Delta T_f$ is the temperature rise of the film. Since the $3\omega$ measurement gives only the temperature difference oscillation amplitude between top (interface heater-sample) and bottom (interface of sample to the infinite sink) of the whole sample, it is not possible to measure $\Delta T_f$ directly. But if one creates a film that consists not only of the film of interest but adds a small reference, the measurable quantity temperature rise of the system ($\Delta T_{sys}$) can be super-posited by the $\Delta T_f$ and the $\Delta T_r$ of the system containing the substrate and the reference (see **Figure S7 *a***). Then the temperature rise of the system can be expressed as:

$$\Delta T_{sys} = \Delta T_r + \Delta T_f \tag{11}$$

The $\Delta T_r$ is obtained directly by measuring other sample that contains only the substrate and the reference film. Finally, the thermal conductivity of the film of interest is obtained by subtracting the $\Delta T_{sys}$ and $\Delta T_f$ and given by:

$$k = \frac{Pd_f}{2bl(\Delta T_{sys} - \Delta T_r)} \tag{12}$$

Therefore, for each film-on-substrate measurement, it is required to create and measure at least two samples, namely one sample containing the film of interest as well as a reference part and a second sample containing only the reference part (see **Figure S7 *c***). Naturally both reference parts need to be created under similar conditions on equal substrates. In our case the reference consisted of 5 nm of vanadium and 25 nm of HfNiSn buffer layers. It is important to mention that as the width of the heater line is not exactly the same in each sample, a correction of the $\Delta T$ must to be applied as follows:

$$\Delta T_c = \frac{b_{mask}}{b_{measured}} \Delta T = c_b \Delta T \tag{13}$$

The correction takes into account the deviation in heater width due to the photolithographic processes. This quantity was estimated through image analysis of five pictures taken with an optical microscope using a 100x objective along of the $3\omega$-strip.

To perform a $T$ measurement with this method, one needs to create a difference of $T$ without a significant $\Delta T$ in the substrate. The error given by this simplification is less than 1%, if the ratio $(k_f/k_S)^2 < 0.01$. In our case all observed $k_f < 5$ W/(K m) and the lowest observed $k_S \approx 50$ W/(K m), then, this criteria is fulfilled for all measurements.[6] The error propagation for the differential method leads to errors about 5%, a mathematical error in this range is tolerable here and the line source criteria does not need to be fulfilled as strictly as for the slope method. To achieve an error lower than 5% the ratio $\lambda/b \leq 2.1$ and for an adequate semi-infinite substrate assumption $d_S/\lambda$ must be bigger than two. The complete requirements for these approximations are shown in **Figure S8**.

### 2.3 Estimation of measurement errors for thin film measurements

For the measurements done in this work, it was always tried to keep the errors caused by the mathematical model as low as possible by choosing an adequate frequency range for present environment temperature. Therefore, the line source assumption and the semi-infinite substrate assumption were taken into account as well as the infinitely long heater assumption. It is not possible to find a range where all limits for an error lower than 1% are fulfilled at the same time for all the temperatures. One reason for this is the fact that the line source criterion behaves in a different way as a function of temperature than the other two criteria. But we can be sure, that the errors caused by mathematical assumptions are always below 5%.

# Supporting Information

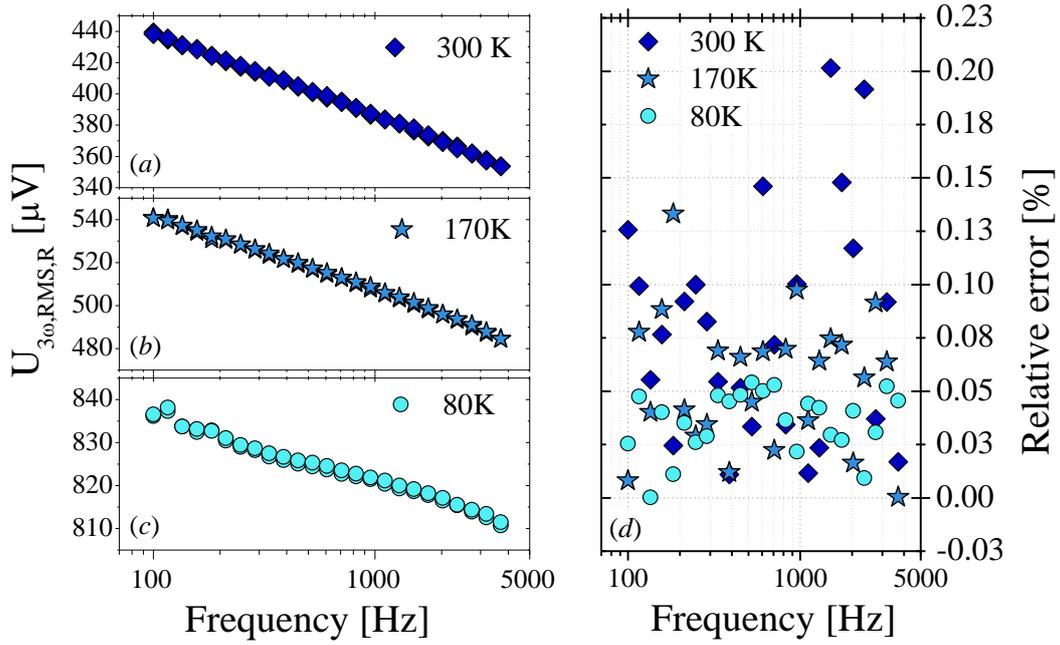

**Figure S9** *Measured three omega voltages (**a**) to (**c**) and relative error (**d**) at each frequency for 1μm thick HfNiSn thin film at three different temperatures T = 300 (**a**), 170 (**b**) and 80 (**c**) K for fixed power P = 20 mW.*

The estimations done by the mathematical model are not the only possible sources of errors. The $3\omega$-method requires several measurements of electrical and geometrical quantities that contain statistical errors that will affect the result as shown in **Table S2** and **Table S3**.

**Table S2** *Relative errors of the measured electrical quantities of the used $3\omega$-method. The errors for $R_0$ and P can be calculated with this values using error propagation.*

| $\delta_{rel}\, dR/dT$ | $\delta_{rel}\, U_{\omega,RMS}$ | $\delta_{rel}\, I_{\omega,RMS}$ | $\delta_{rel}\, U_{3\omega,RMS}$ |
|---|---|---|---|
| 0.5 % | 0.1 % | 0.1 % | 0.3 % |

In contrast to the other measured quantities, the determination of the third omega signal $U_{3\omega,\,RMS}$ is not straightforward. Therefore one needs to explain how the errors are determined in this case. For each sample at each measured temperature, the chosen frequency range is measured at least twice. The error was determined by the deviation of both measured points. An example for this praxis is showed in **Figure S9**. The deviation of a point at each frequency leads to relative errors less than 0.2% in this example. The highest relative error of $U_{3\omega,\,RMS}$ measured in this work was 0.3%.

**Table S3** *Relative errors of the measured geometrical quantities of the used $3\omega$-method.*

| $\delta_{rel}\, d_f$ | $\delta_{rel}\, l$ | $\delta_{rel}\, b$ | $\delta_{rel}\, c_b$ |
|---|---|---|---|
| 2.0 % | 0.5 % | Expressed in $c_b$ | 1.0 % |

The propagation of uncertainty is done below for the absolute error $\delta x$ of the general quantity $x$. The single quantities are assumed as uncorrelated. As an example, the error propagation for the temperature coefficient of the resistance $\beta$ is calculated below:

$$\delta \beta = \sqrt{\left(\frac{\delta dR/dT}{R_0}\right)^2 + \left(\frac{\delta R_0 dR/dT}{R_0^2}\right)^2} \tag{14}$$

The error propagation for the corrected temperature oscillation amplitude $\delta \Delta T_c$ can be written as:

$$\delta \Delta T_c = 4 \left[ \frac{c_b^2 \delta U_{\omega,RMS}^2 U_{3\omega,RMS}^2}{\beta^2 U_{\omega,RMS}^4} + \frac{c_b^2 \delta \beta^2 U_{3\omega,RMS}^2}{\beta^4 U_{\omega,RMS}^2} \right. \\ \left. + \frac{c_b^2 \delta U_{\omega,RMS}^2}{\beta^2 U_{\omega,RMS}^2} + \frac{\delta c_b^2 U_{3\omega,RMS}^2}{\beta^2 U_{\omega,RMS}^2} \right]^{1/2} \tag{15}$$

The error propagation for the $k$ of the differential method can be expressed as:

$$\delta k = \left[ \frac{d_f^2 \delta l^2 P^2}{4b^2 l^4 (\Delta T_{sys,c} - \Delta T_{r,c})^2} + \frac{d_f^2 \delta \Delta T_{r,c}^2 P^2}{4b^2 l^2 (\Delta T_{sys,c} - \Delta T_{r,c})^4} + \frac{d_f^2 \delta \Delta T_{sys,c}^2 P^2}{4b^2 l^2 (\Delta T_{sys,c} - \Delta T_{r,c})^4} \right. \\ \left. + \frac{d_f^2 \delta P^2}{4b^2 l^2 (\Delta T_{sys,c} - \Delta T_{r,c})^4} + \frac{\delta d_f^2 P^2}{4b^2 l^2 (\Delta T_{sys,c} - \Delta T_{r,c})^2} \right]^{1/2} \tag{16}$$

Due to the mathematical expressions for $\beta$ and $\Delta T_c$ only contain products, the relative errors $\delta_{rel} x = \delta x / x$ can be expressed as:

$$\delta_{rel}\beta = \frac{\delta \beta}{\beta} = \sqrt{(\delta_{rel} dR/dT)^2 + (\delta_{rel} R_0)^2} \tag{17}$$

$$\delta_{rel}\Delta T_c = \frac{\delta \Delta T_c}{\Delta T_c} = \sqrt{(\delta_{rel} c_b)^2 + (\delta_{rel} U_{\omega,rms})^2 + (\delta_{rel} U_{3\omega,rms})^2 + (\delta_{rel}\beta)^2} \tag{18}$$

One sees here, that the relative error of $\Delta T_c$ depends only on the relative errors of measurement instruments. Usually, it is possible to assume that the relative error of an instrument only dependents on the chosen measurement scale. Then, if one uses the same scale for each measured sample, the relative error does neither depend on the single measured value of a quantity nor on the measured sample. Therefore, we can assume that the relative error $\delta_{rel}\Delta T_c$ is constant for each measurement. The case is different for the $k$, because, $k$ contains the difference $(\Delta T_{sys,c} - \Delta T_{r,c})$, thus it cannot be expressed just in terms of relative errors and therefore does not stay constant. But, the largest part of it can be expressed with relative measurement errors, so that it shows only a dependence on the ratio $\Delta T_{sys,c}/\Delta T_{r,c}$:

# Supporting Information

$$\delta_{rel}k = \frac{\delta k}{k} = \left[(\delta_{rel}d_f)^2 + (\delta_{rel}l)^2 + (\delta_{rel}P)^2 \right.$$
$$\left. + (\delta_{rel}\Delta T_{r,c})^2(\Delta T_{sys,c}/\Delta T_{r,c} - 1)^{-2} + (\delta_{rel}\Delta T_{sys,c})^2(1 - (\Delta T_{sys,c}/\Delta T_{r,c})^{-1})^{-2}\right]^{1/2} \quad (19)$$

In **Figure S10** this function is plotted for the instrument errors shown in **Table S2** and **Table S3**. Notice that the relative error of $k$ decreases with increasing $\Delta T_{sys,c}/\Delta T_{r,c}$. Therefore, it is advisable to grow samples with a thickness difference between reference and film of interest as large as possible. In this way it is warranted that the $\Delta T$ ratio is large enough. It is important to notice that the $\Delta T_{sys,c}/\Delta T_{r,c}$ is also dependent on the difference of $k$ between reference and film of interest, then, it will not be similar for samples with equal thickness. In our work the $\Delta T_{sys,c}/\Delta T_{r,c}$ is close to 1.5 for most of the samples measured in this work. This leads to errors between 4 and 7%.

For all the measurements carried out in this work, at least two measurement points of $U_{3\omega,rms}$ were taken for similar conditions of the same sample (See **Figure S7 c**). In this way, one can see statistical deviations as well as the errors calculated with error propagation. This statistical deviations are much smaller than the error bars calculated with the Eq. (19), because they only depend on deviations of $U_{3\omega,rms}$ while the error propagation takes several additional error sources into account. Therefore the error propagation is more feasible than considering simply the statistical errors of $U_{3\omega,rms}$.

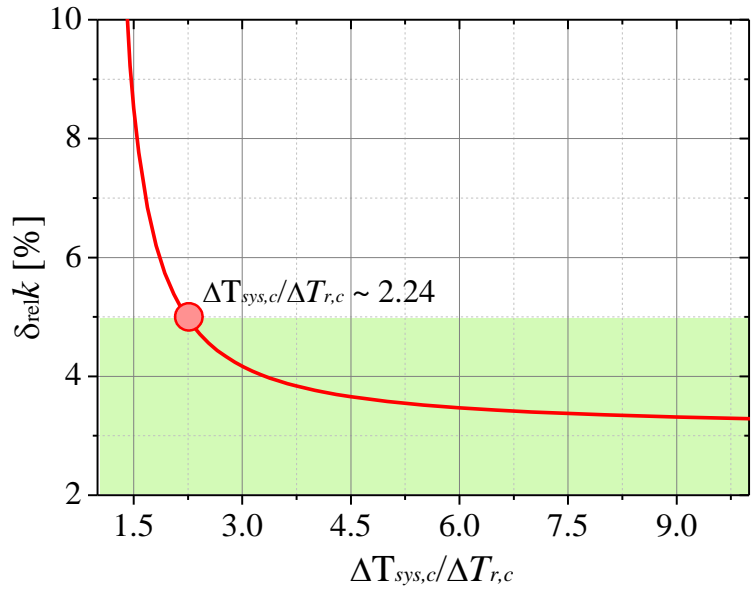

**Figure S10** *Relative error of the thermal conductivity for the differential method in dependence of the ratio $\Delta T_{sys,c}/\Delta T_{r,c}$. The curve is dependent on the relative errors of the measured quantities. This curve was calculated for the estimated errors of the 3ω setup used in this work.*

# Supporting Information

## 3 Calculation of amorphous limit of the thermal conductivity

The lowest thermal conductivity value for semiconductors and insulating materials is achieved for the systems with the small order what is similar to the amorphous state. The pioneer theoretical framework on heat conduction in amorphous materials was first proposed by Einstein[10], refined by Slack[11] and extended by Cahill et al.[12] The theory is basically based on the assumption of that heat conduction is described by a "random walk" of independent oscillators with a characteristic frequency (Einstein frequency). Then, each atom is coupled to its first-, second-, and third nearest neighbors on a simple cubic lattice by harmonic forces. Slack reformulated this problem by considering that the minimum MFP ($\Lambda$) of a heat carrier has to be the same as its wavelength ($\lambda$), namely $\Lambda = \lambda$.[11] The $k$ estimated by this model is known as the minimum thermal conductivity or amorphous limit ($k_{min}$). Following both works, Cahill et al.[12] further extended this model by dividing the system into regions of size $\lambda/2$, with a constant velocity given by the Debye speed of sound. Then, the $\Lambda$ of each oscillator is assumed to be $\lambda/2$. Finally, the $k$ is reformulated in terms of sum of three Debye integrals as follows[12]:

$$k_{min} = \left(\frac{\pi}{6}\right)^{1/3} k_B n^{2/3} \sum_j v_j \left(\frac{T}{\Theta_{D,j}}\right)^2 \int_0^{\Theta_{D,j}/T} \frac{x^3 e^x}{(e^x - 1)^2} dx \qquad (20)$$

where $\Sigma_j$ represents the sum on longitudinal ($L$) and transverse ($T$) polarizations, $n$ is the number density of atoms (i.e., $n$ = number of atoms in a unit cell / volume of unit cell), $k_B$ is the Boltzmann constant, $\Theta_{D,i}$ is the Debye temperature given by:

$$\Theta_{D,i} = \frac{h v_i}{2\pi k_B} (6\pi^2 n)^{1/3} \qquad (21)$$

where $h$ is the Planck constant. The **Table S4** summarizes all the parameters used in this work to calculate the amorphous limit of the HH compounds.

**Table S4** *Parameter used to estimate the amorphous limit of the thermal conductivity of the HHs compounds.*

| Parameter | TiNiSn | HfNiSn |
|---|---|---|
| $v_L$ | 5952 [m/s][13] | 4195 [m/s][14] |
| $v_T$ | 3427 [m/s][13] | 2783 [m/s][14] |
| $a$ | 0.5941 [nm][15] | 0.6083 [nm][16] |
| *Number of atoms per unit cell* | 12 | 12 |

# Supporting Information

The amorphous limit of the effective material was calculated as the reciprocal of the average of the minimum thermal conductivities. The temperature dependence of the amorphous limit for each compound is displayed in **Figure S11**.

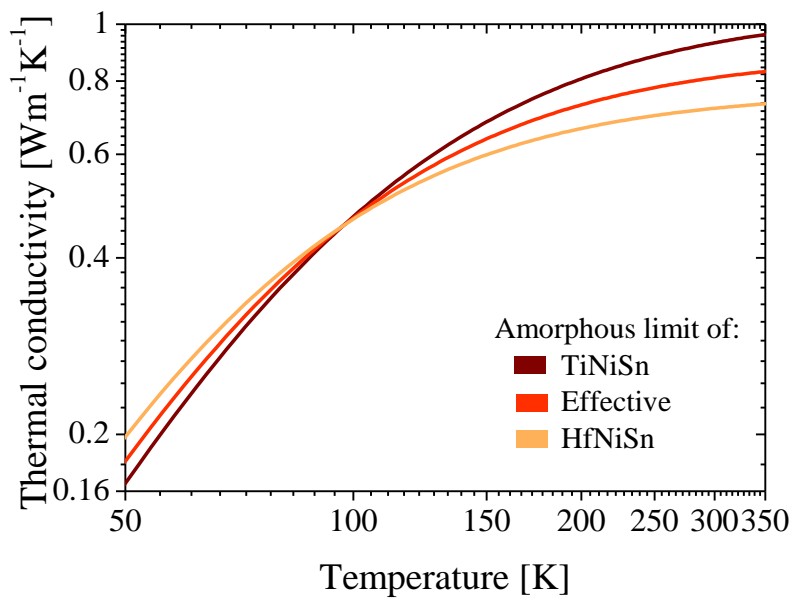

**Figure S11** *Amorphous limit of the different HH compounds and the effective material.*

**References**

(1) Komar, P.; Jakob, G. CADEM : Calculate X-Ray Diffraction of Epitaxial Multilayers. *J. Appl. Crystallogr.* **2017**, *50* (1), 288–292.

(2) Cahill, D. G. Thermal Conductivity Measurement from 30 to 750 K: The 3ω Method. *Rev. Sci. Instrum.* **1990**, *61* (2), 802.

(3) Cahill, D. G. Erratum: "Thermal Conductivity Measurement from 30 to 750 K: The 3ω Method" [Rev. Sci. Instrum. 61, 802 (1990)]. *Rev. Sci. Instrum.* **2002**, *73* (10), 3701.

(4) Carslaw, H. S. .; Jaeger, J. C. . The Flow of the Heat in an Infinite Circular Cylinder. In *Conduction of Heat in Solids*; Oxford University Press: London, 1959; pp 188–214.

(5) Borca-Tasciuc, T.; Kumar, A. R.; Chen, G. Data Reduction in 3ω Method for Thin-Film Thermal Conductivity Determination. *Rev. Sci. Instrum.* **2001**, *72* (4), 2139–2147.

(6) Dames, C. Measuring the Thermal Conductivity of Thin Films: 3 Omega and Related Electrothermal Methods. *Annu. Rev. Heat Transf.* **2013**, *16* (1), 7–49.

(7) Cahill, D. G. . Thermal conductivity bulk single crystal MgO; Temperature dependence of volumetric heat capacity MgO. Department of Materials Science and Engineering. University of Illinois at Urbana Campaign http://users.mrl.illinois.edu/cahill/tcdata/tcdata.html (accessed Jan 1, 2016).

(8) Cahill, D.; Katiyar, M.; Abelson, J. Thermal Conductivity of a-Si:H Thin Films. *Phys. Rev. B* **1994**, *50* (9), 6077–6081.

(9) Kim, J. H.; Feldman, A.; Novotny, D. Application of the Three Omega Thermal Conductivity Measurement Method to a Film on a Substrate of Finite Thickness. *J. Appl. Phys.* **1999**, *86* (7), 3959–3963.

(10) Einstein, A. Elementare Betrachtungen Über Die Thermische Molekularbewegung in Festen Körpern. *Ann. Phys.* **1911**, *340* (9), 679–694.

(11) Slack, G. A. The Thermal Conductivity of Nonmetallic Crystals. In *Solid State Physics Vol. 34*; Ehrenreich, H., Seitz, F., Turnbull, D., Eds.; Academic Press: New York, San Francisco & London, 1979; pp 1–71.

(12) Cahill, D. G.; Watson, S. K.; Pohl, R. O. Lower Limit to the Thermal Conductivity of Disordered Crystals. *Phys. Rev. B* **1992**, *46* (10), 6131–6140.

(13) Hichour, M.; Rached, D.; Khenata, R.; Rabah, M.; Merabet, M.; Reshak, A. H.; Bin Omran, S.; Ahmed, R. Theoretical Investigations of NiTiSn and CoVSn Compounds. *J. Phys. Chem. Solids* **2012**, *73* (8), 975–981.

(14) Özisik, H.; Çolakoglu, K.; Özisik, H. B. Ab-Initio First Principles Calculations on Half-Heusler NiYSn (Y=Zr, Hf) Compounds Structural, Lattice Dynamical and Thermodynamical Properties. *AJP Fiz.* **2010**, *16* (2), 154–157.

(15) Romaka, V. V.; Rogl, P.; Romaka, L.; Stadnyk, Y.; Melnychenko, N.; Grytsiv, A.; Falmbigl, M.; Skryabina, N. Phase Equilibria, Formation, Crystal and Electronic Structure of Ternary Compounds in Ti–Ni–Sn and Ti–Ni–Sb Ternary Systems. *J. Solid State Chem.* **2013**, *197*, 103–112.

(16) Larson, P.; Mahanti, S. D.; Kanatzidis, M. G. Structural Stability of Ni-Containing Half-Heusler Compounds. *Phys. Rev. B* **2000**, *62* (19), 12754–12762.